\documentclass[%
 reprint,
 amsmath,amssymb,
 aps,
]{revtex4-2}

\usepackage{graphicx}
\usepackage{dcolumn}
\usepackage{bm}

\usepackage{xcolor}
\usepackage{hyperref}

\begin{document}

\preprint{APS/123-QED}

\title{Social hierarchy shapes foraging decisions}

\author{Lisa Blum Moyse}
\author{Ahmed El Hady}%
 \altaffiliation[Also at ]{Department of Collective Behavior, Max Planck Institute of Animal Behavior, Konstanz, Germany} 
\affiliation{%
Centre for the Advanced Study of Collective Behaviour,  University of Konstanz, Konstanz, Germany 
}%




\date{\today}

\begin{abstract}
Social foraging is a widespread form of animal foraging in which groups of individuals coordinate their decisions to exploit resources in the environment. Animals show a variety of social structures from egalitarian to hierarchical. In this study, we examine how different forms of social hierarchy shape foraging decisions. We developed a mechanistic analytically tractable model to study the underlying processes of social foraging, tying the microscopic individual to the macroscopic group levels. Based on a stochastic evidence accumulation framework, we developed a model of patch-leaving decisions in a large hierarchical group with leading and following individuals.  Across a variety of information sharing mechanisms, we were able to analytically quantify emergent collective dynamics.  We found that follower-leader dynamics through observations of leader movements or through counting the number of individuals in a patch confers, for most conditions, a benefit for the following individuals by increasing their accuracy in inferring patch richness. On the other hand, misinformation, through the communication of false beliefs about food rewards or patch quality, shows to be detrimental to following individuals, but paradoxically may lead to increased group cohesion. In an era where there is a huge amount of animal foraging data collected, our model provides a systematic way to conceptualize and understand those data by uncovering hidden mechanisms underlying social foraging decisions.


\end{abstract}

\maketitle

\makeatletter
\def\l@paragraph{\@dottedtocline{5}{5.3em}{2.1em}}
\makeatother

 

\paragraph*{Introduction---}Foraging is a fundamental decision-making behavior crucial for animal survival~\cite{stephensForagingBehaviorEcology2007}. Often, animals forage socially and adopt a wide variety of social organizations, from the less to the more hierarchical. Some examples of animals adopting hierarchical structures are eusocial insects~\cite{detrainCollectiveDecisionMakingForaging2008}, birds~\cite{bakerForagingSuccessJunco1981}, mice~\cite{leeForagingDynamicsAre2018} or baboons~\cite{strandburg-peshkinSharedDecisionmakingDrives2015}. While social hierarchy shapes a wide variety of behavioral facets~\cite{tibbettsEstablishmentMaintenanceDominance2022a}, interactions with social foraging are not well understood in quantitative mechanistic terms. To investigate the underlying processes of social foraging, several theories and quantitative models have been developed, such as the ideal free distribution~\cite{fretwellTerritorialBehaviorOther1969}, the marginal value theorem for groups~\cite{livoreilPatchDepartureDecisions1997}, agent-based models~\cite{wajnbergOptimalPatchMovement2013}, reinforcement learning models~\cite{falcon-cortesCollectiveLearningIndividual2019}\cite{lofflerCollectiveForagingActive2023}, game-theoretic approaches~\cite{giraldeauSocialForagingTheory2000}\cite{cressmanGameTheoreticMethodsFunctional2014}, or Bayesian models~\cite{perez-escuderoCollectiveAnimalBehavior2011}. However, these approaches are often coarse-grained, implying general interaction rules, or are too complex to be treated analytically. Another limitation is that uncertainty and noise in decision processes are often not taken into account, which can strongly shape the agents' patch departure statistics while foraging~\cite{davidsonForagingEvidenceAccumulation2019}\cite{kilpatrickUncertaintyDrivesDeviations2021a}. This situation calls for the development of an analytically tractable framework, including inherent stochasticity, to study the mechanisms underlying social foraging across a variety of social and environmental conditions. Models developed within this framework should also be able to be fitted to experimental data, to unravel mechanisms underlying the decision-making of animals while they forage socially. In this letter, we introduce a new framework to meet these expectations.\\
The new model developed in this paper is built on a widespread stochastic decision-making framework based on the evidence accumulation process, 
drift-diffusion models, which can be applied to patch-leaving tasks~\cite{davidsonForagingEvidenceAccumulation2019}.
So far, such mechanistic models have focused only on one~\cite{davidsonForagingEvidenceAccumulation2019} or two cooperative foragers~\cite{bidariStochasticDynamicsSocial2022}. We have previously extended this modeling approach to study coupled agents in an egalitarian group for a two-patches environment~\cite{blummoyseSocialPatchForaging2024}. 
\begin{figure}   \includegraphics[width=0.4\textwidth]{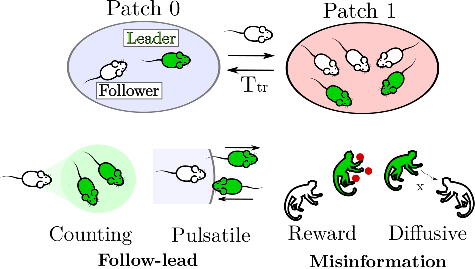}
    \caption{(Top) Schema of a hierarchical foraging group in a two-patches environment, with a travel time $T_\text{tr}$. (Bottom) Followers collect social information from leaders in different ways, to follow leaders  (counting, pulsatile) or be misinformed about patch qualities  (reward, diffusive).}
    \label{fig:intro}
\end{figure}

In this letter, we provide a quantitative analytically tractable model to unravel the mechanisms of how social hierarchy shapes social foraging decisions strategies. For simplicity, we focus on a hierarchical organization in a two-patches environment with leading and following agents. We analytically derive strategies under a variety of information sharing mechanisms, gathered in two categories, follow-lead and misinformation (see Fig.~\ref{fig:intro}). The second category, also known as tactical deception, is more likely to occur mainly in primates~\cite{mitchellDeceptionPerspectivesHuman1986}. Throughout this letter, numerical simulations and analytical analysis enable us to characterize how these different information sharing mechanisms lead to various collective dynamics. We found that each information sharing mechanism shapes decision strategies in a specific way. For example, misinformation through false belief sharing decreases patch richness accuracy estimation, while increasing group cohesion. We think that this model can provide a benchmark for evolutionary comparison across species, opening up the ability for quantitative behavioral evolutionary studies.\\


\paragraph*{Social foraging models---}

\begin{figure}   \includegraphics[width=0.45\textwidth]{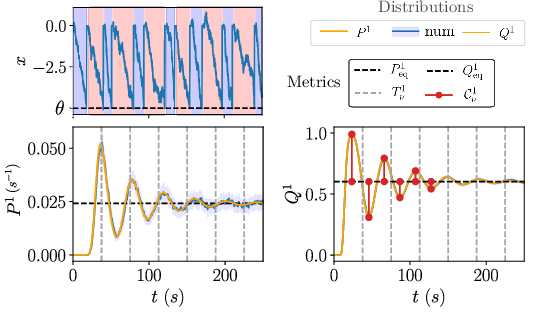}
    \caption{Dynamics and quantification of the collective foraging task. (Top) Foraging agents accumulate evidence in a patch of food (blue: patch 0, red: patch 1), and leave it  for the next patch when their decision variable x reaches the threshold $\theta$. (Bottom) Distributions of leaving times $P^1(t)$ (Left) and fraction of agents $Q^1(t)$ (Right) in patch 1 show damped oscillations converging to an equilibrium state. Blue lines correspond to numerical simulations, and orange lines to theoretical predictions. The shaded area corresponds to standard deviation. The distributions can be quantified by their equilibrium values, patch residence time and measure of cohesion. Parameters are $\theta=-5$ and $\alpha=1$~s$^{-1}$, $B = 0.05$~s$^{-1}$, $p^0=0.6$, $p^1=0.8$, $\tau_d=0.1$~s, $T_\text{tr}=2$~s, $N_L=2500$, $20$ simulations.}
    \label{fig:introb}
\end{figure}

Agents move between two patches identified by a number $k$. $k=0$ is the initial patch, and $k=1$ is the highest quality patch. The decision to leave a patch for another one is taken through the underlying process of evidence accumulation.
Groups are organized hierarchically, with a subgroup of leading (L) and following (F) agents. Leading individuals do not consider the information coming from other agents, they act like non-interacting agents. Following individuals have non-zero coupling parameters, and only follow leading individuals. In addition, they receive less food.\\

$x_i$, the decision variable of a foraging agent $i$ in a patch $k$, evolves according to the following stochastic differential equation

\begin{equation}
dx_i(t)=(r_i(t)-\alpha)dt + \sum_{j \in L} c_{ij}(t-\tau_d)dt+\sqrt{2B} dW_i(t)
\label{x}
\end{equation}

with the initial condition $x_i(0)=0$ for all $i$~s. The forager leaves a patch when $x_i$ reaches the threshold $\theta$, and $x_i$ is reset to zero. 
The travel time between patches is $T_\text{tr}$. The drift-diffusion process is illustrated in Fig.~\ref{fig:introb}.\\
The decision variable evolves following an evidence accumulation process, where $\alpha$ represents the cost associated with foraging, and $r_i(t)$ the food rewards.
Every time step $\Delta_{r}$, an L agent has a probability $p^{k}$ of receiving a food reward $r_i=1$~s$^{-1}$. Throughout this letter, we fix $\Delta_{r}=dt$. The following individuals have a reward probability $\beta p^k$, with $\beta\leq1$. $p^k$ is constant here, this is a non-depleting environment. $W_i(t)$ is the standard Wiener process, $B$ represents the noise amplitude. For F agents, information sharing from a leading $j$ to a following $i$ individual in the same patch is expressed by the coupling term $c_{ij}$, with a delay $\tau_d$. For L agents, this term is equal to zero. The different coupling types are illustrated in Fig.~\ref{fig:intro} and detailed below.\\

\textit{Follow-lead dynamics.} Two different mechanisms may underlie the following dynamics.
First, F agents could perceive the beliefs of L agents through pulses of information, corresponding to an agent's arrival ($\kappa_a$) or departure ($\kappa_d$) in a patch. This is the \textit{pulsatile coupling}. If an L agent $j$ leaves, $c_{ij}(t)= -{\kappa_d}{(N_L dt)^{-1}}\delta(x_j-\theta)$. If an L agent $j$ leaves to join the current patch of a F agent $i$, $c_{ij}(t)= {\kappa_a}{(N_L dt)^{-1}} \delta(x_j-\theta)$.\\
Instead of being sensitive to the departure or arrival times of L individuals, F agents could also perceive how many L individuals of their group are in the same patch. This is the \textit{counting coupling}. For clarity reasons, we write this term without the sum in eq.~\eqref{x},
$c_{i}(t) = \kappa_c (n^{k}_L(t)/N_L-\eta)$.
If many L agents are in the same patch, $n_L^{k}(t)>\eta$, so $c_{i}>0$, i.e. the F agent accumulates evidence to stay in the patch. Inversely for $n_L^{k}(t)<\eta$.\\

\textit{Misinformation} L agents may communicate false information to mislead F individuals. L agents may continuously share their beliefs about patch quality, this is the \textit{diffusive coupling}. $c_{ij}(t) = \kappa_\text{diff}^k {N_L^{-1}} x_{L,j}(t)$. Lying L agents may modify their communicated decision variable, to misinform F agents. If $\kappa_\text{diff}^1\geq\kappa_\text{diff}^0$, this incites F agents to leave the best patch.\\
False information can also come from a wrong communicated number of catches that L agents in the same patch get. $c_{ij}(t) = {\kappa_r}{N_L^{-1}} r_j(t)$. 
$\kappa_r$ corresponds to the \textit{reward coupling} strength. The false communicated reward probability is
$\zeta^k$. If $\zeta^1\leq\zeta^0$, this incites F agents to stay in the worst patch.\\
Pulsatile, diffusive and reward coupling results with normalization terms $n^{k}_L(t)$ instead of $N_L$ are presented in the appendix.\\

\begin{figure*}
    \centering
\includegraphics[width=0.99\linewidth]{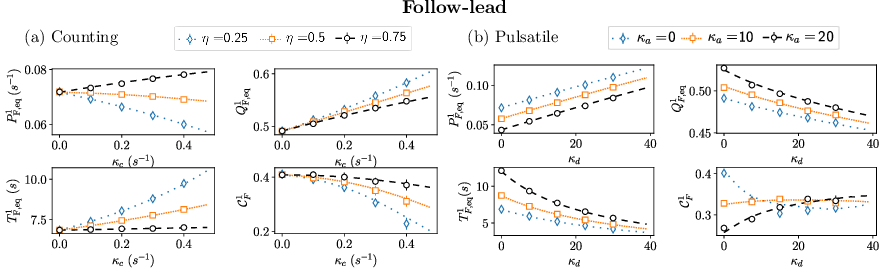}
    \caption{Follow-lead dynamics of the best patch metrics: equilibrium leaving time distribution $P_{F,\text{eq}}^1$, equilibrium accuracy $Q_{F,\text{eq}}^1$, equilibrium patch residence time $T_{F,\text{eq}}^1$, and cohesion $\mathcal{C}_F^1$. (a) Counting coupling metrics as a function of the coupling strength parameter $\kappa_c$ for different $\eta$ values. (b) Pulsatile coupling metrics as a function of the departure strength parameter $\kappa_d$ for different arrival strength parameter $\kappa_a$ values. Lines correspond to theoretical predictions and dots to numerical simulations. Error bars correspond to the standard deviation.}
    \label{FL}
\end{figure*}

\textit{Simulations details.}
The stochastic differential equation~\eqref{x} of the decision variable is solved numerically with the Euler method. The parameters $\theta=-5$, $\alpha=1$~s$^{-1}$, $B = 0.05$~s$^{-1}$, $p^0=0.6$, $p^1=0.9$, $\tau_d=0.1$~s, $T_\text{tr}=0.5$~s, $\beta = 0.3$ remain constant throughout the figures. The coupling parameters remain small enough compared to the asocial drift term. The duration of the simulation is equal to $250$~s. Cohesion is quantified on the $\epsilon=6$ first extrema. The L and F group sizes are $N_L=200$ and $N_F=200$, with $dt=0.001$~s, $20$ simulations, for all couplings except for pulsatile, with $N_L=60000$ and $N_F=2000$, 2 simulations. A larger L group size was needed for this condition, to get a correct match between theory and simulations. A larger time step $dt = 0.02$~s was implemented to keep tractable computation times. In fact, leaving probability densities $P_L^k(t)$ are included in the drift terms of pulsatile coupling, and they are more noisy than the fraction of agents $Q_L^k(t)$.  Equilibrium metrics from numerical simulations were computed through the averaged last 100~s of the process.\\
The distributions and metrics used to quantify the process are first described, and then applied to the follow-lead and misinformation dynamics.\\





\paragraph*{Distributions and metrics---}
The probability density of a group $X$ ($X=\{L,F\}$) to leave a patch $k$ at time $t$ is given by\\
$P_X^k(t) = \sum_{\nu=1}^\infty P_{X,\nu}^k\bigl(t-(2\nu-2+k) T_\text{tr}\bigr)$~\cite{blummoyseSocialPatchForaging2024}, with
\begin{equation}
  P_{X,\nu}^k(t) = \underbrace{(\Psi^k_{X,1}* ...*\Psi^0_{X,1}*\Psi^1_{X,1}*\Psi^0_{X,1} )}_{2\nu-1+k}(t)
\end{equation}
For $t\leq (2\nu-2+k)T_\text{tr}$, the convolution is equal to zero.
With $\Psi^k_{X,\nu}(t)=\frac{-\nu\theta}{\sqrt{4\pi B t^3}} \exp \biggl(\frac{-(\nu\theta+\widetilde{\alpha}_X^k t)^2}{4Bt} \biggr)$~\cite{coxTheoryStochasticProcesses1977} for $\nu\geq 1$, and $\Psi^k_{X,0}(t) = 1$ for all $t$. $\widetilde{\alpha}_X^k$ is the effective drift term. 

For L agents, the effective drift term is $\widetilde{\alpha}_L^k = \alpha-p^k$. The mean rate is equal to $p^k$ (with unit $s^{-1}$). For clarity reasons, we write directly $p^k$ in effective drifts. Since it is constant over time, the L leaving probability density can be found after using the Laplace transform of $\Psi^k_{X,1}$s to compute the convolution of functions:
$P_L^k(t) = \sum_{\nu=1}^\infty \bigl(\Psi^0_{L,\nu}*\Psi^1_{L,\nu-1+k}\bigr)\bigl(t-(2\nu-2+k) T_\text{tr}\bigr)$.\\
F agents dynamics are described below.\\

The probability to be in the patch at a time $t$ is given by the time integral of the arrival minus the departure probability density in a patch $k$:
\begin{equation}
    Q_X^k(t) = 1-k + \int_0^t d\tau [P_X^{k'}(\tau-T_\text{tr})-P_X^k(\tau)] 
    \label{Qk}
\end{equation}
Throughout this letter, we refer to the fraction of agents in the best patch $Q_X^1(t)$ as accuracy.\\
Fig.~\ref{fig:introb} shows an example of distributions $P^k_L(t)$, $Q_L^k(t)$. After an oscillating period, they converge towards an equilibrium value. These equilibrium values can be calculated~\cite{blummoyseSocialPatchForaging2024}: ${T}_{X,\text{eq}}^k= -\frac{\theta}{\widetilde{\alpha}_X^k}$, $\ Q_{X,\text{eq}}^k = \biggl(-\frac{2\widetilde{\alpha}_X^k T_\text{tr}}{\theta} + 1 + \frac{\widetilde{\alpha}_X^k}{\widetilde{\alpha}_X^{k'}}\biggr)^{-1}$\\ $P_{X,\text{eq}}^k =\widetilde{\alpha}_X^k\Biggl(2\widetilde{\alpha}_X^k T_\text{tr} - \theta\biggl(1 + \frac{\widetilde{\alpha}_X^k}{\widetilde{\alpha}_X^{k'}}\biggr)\Biggr)^{-1}$\\

Cohesion is quantified as the deviation of extrema from equilibrium. For $\epsilon$ local maxima and minima $Q^k_{X,\nu}$, the cohesion metrics is defined as $\mathcal{C}_X^k =\frac{1}{\epsilon}\sum_{\nu=1}^\epsilon Q^k_{X,\nu}  - Q_{X,\text{eq}}^k$.
The larger $\mathcal{C}_X^k$ is, the less damped the oscillations are, i.e. the more cohesive the group is.\\

\begin{figure*}
    \centering
\includegraphics[width=0.99\linewidth]{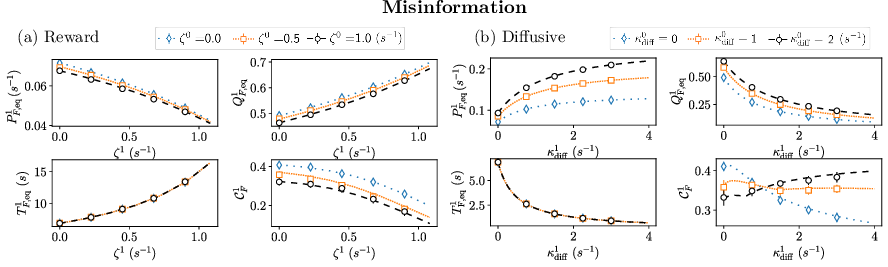}
    \caption{Misinformation dynamics of the best patch metrics: equilibrium leaving time distribution $P_{F,\text{eq}}^1$, equilibrium accuracy $Q_{F,\text{eq}}^1$, equilibrium patch residence time $T_{F,\text{eq}}^1$, and cohesion $\mathcal{C}_F^1$. (a) Reward coupling metrics as a function of the communicated reward rates in patch 1 $\zeta^1$, for different communicated reward rates in patch 0 $\zeta^0$ values. The reward strength parameter is $\kappa_r=0.5$. (b) 
    Diffusive coupling metrics as a function of the diffusive strength parameter in patch 1 $\kappa_\text{diff}^1$, for different diffusive strength parameter in patch 0 $\kappa_\text{diff}^0$ values. Lines correspond to theoretical predictions and dots to numerical simulations. Error bars correspond to the standard deviation.}
    \label{Lie}
\end{figure*}

\paragraph*{Follow-Lead dynamics---}

To analytically predict the F distributions and metrics, it is possible to include social information in an effective drift term 
$\widetilde{\alpha}_F^k$. If the L dynamics is slow enough compared to the F one and $N_L>>1$ (see supplementary Fig.~\ref{SuppFig}(a) for theory-simulation fit as a function of $N_L$), the averaged term over the time spent in a patch $k$, $\bigl<\widetilde{\alpha}_F^k\bigr>(t,\tau)$, may be used as a quasi-continuous drift term in the distribution calculations.
$\bigl<\widetilde{\alpha}_F^k\bigr>(t,\tau) = {(t-\tau-T_\text{tr})^{-1}}\int_{\tau+T_\text{tr}}^t \widetilde{\alpha}_F^k(\tau')d\tau'$ 
with the arrival ($\tau$) and departure ($t$) times. At equilibrium, the coupling parameters are restricted by $\alpha_{F,\text{eq}}^k\geq 0$~s$^{-1}$.\\

For counting coupling,
\begin{equation}
    \widetilde{\alpha}_F^k(t) = \alpha - \beta p^k - \kappa_c (Q^k_L(t)-\eta)
\end{equation}

Fig.~\ref{FL}(a)
shows that an increasing coupling parameter $\kappa_c$ is associated with an increased equilibrium accuracy, patch residence time, and decreased cohesion. The effect is more important for small $\eta$ values. The equilibrium leaving probability density increases with $\kappa_c$ for large $\eta$ values and decreases for small ones.\\

The effective drift term with pulsatile coupling depends on the departure and arrival probability densities.
\begin{equation}
    \widetilde{\alpha}_F^k(t) = \alpha - \beta p^k + \kappa_d P^k_L(t) - \kappa_a P^{k'}_L(t)
\end{equation}






Fig.~\ref{FL}(b)
shows that higher $\kappa_d$ values are associated with a decrease in equilibrium accuracy, patch residence time, and an increased equilibrium leaving probability density. These effects are more important for small $\kappa_a$ values. Please note that in the case of normalization by $n_L^k$ instead of $N_L$, increased $\kappa_d$ values are linked to an increased accuracy, see supplementary Fig.~\ref{SuppFig}(b). Cohesion decreases for small $\kappa_a$ values, and increases for large ones. All in all, departure and arrival coupling have opposite effects on accuracy, patch residence time and leaving probability, while their impact on cohesion is less trivial and depends on $\{\kappa_d$, $\kappa_a\}$ combinations. Cohesion decreases with increasing departure strengths $\kappa_d$ for small $\kappa_a$ values, and increases for larger $\kappa_a$ values.\\

\paragraph*{Lying: impact of misinformation---}


In this section, lying L agents display or communicate wrong information about the environment richness (reward coupling) or their belief about a patch quality (diffusive coupling).\\
To communicate wrong social information about patch quality, lying agents increase their communicated number of catches in the worst patch ($k=0$) and decrease it in the best patch ($k=1$). In this way, F agents are misinformed about the distribution of resources and perceive false reward rates $\zeta^k$.
The effective drift term is
\begin{equation}
    \widetilde{\alpha}_F^k(t)  = \alpha - \beta p^k -\kappa_r \zeta^k Q^k_L(t)
\end{equation}

Fig.~\ref{Lie}(a) shows the impact of reward misinformation.
A smaller $\zeta^1$ value is associated with a decrease in equilibrium accuracy, patch residence time, and an increase in leaving density and cohesion. Larger $\zeta^0$ values contribute to the decrease in accuracy, leaving density and cohesion, while having no impact on patch residence time.\\

To communicate wrong beliefs about patch quality, lying agents increase their decision variable in the worst patch ($k=0$) and decrease it in the best patch ($k=1$). In this way, F agents are misinformed about the leading agents' beliefs and perceive modified averaged decision variables $\kappa_\text{diff}^k \bigl<x_L^k\bigr>(t)$. The effective drift term is
\begin{equation}
    \widetilde{\alpha}^k_F(t) = \alpha - \beta p^k - \kappa_\text{diff}^k \bigl<x_L^k\bigr>(t)
\end{equation}
See the appendix for $\bigl<x_L^k\bigr>(t)$ calculation details.\\
At equilibrium, the averaged L decision variable can be estimated as $x^k_\text{L,eq} = Q_{L,\text{eq}}^k\biggl(\frac{\theta}{2} + \frac{B}{\widetilde{\alpha}_L^k} \biggr)$. The first part corresponds to the mean and the second part represents the variance of the Ornstein-Uhlenbeck process~\cite{gardinerHandbookStochasticMethods2004}.\\
Fig.~\ref{Lie}(b) shows the impact of misinformation on patch quality belief. A larger $\kappa_\text{diff}^1$ is associated with a decrease in accuracy,  patch residence time, and an increased leaving density. Smaller $\kappa_\text{diff}^0$ values contribute to the decrease in accuracy, leaving density, while does not have an impact on patch residence time. The cohesion dynamics is more complex and depends on the $\{\kappa_\text{diff}^0,\kappa_\text{diff}^1\}$ combinations. Larger $\kappa_\text{diff}^1$ values decrease accuracy for small $\kappa_\text{diff}^0$ ones, and inversely for large $\kappa_\text{diff}^0$ values.\\






\paragraph*{Discussion---} 
This letter introduces a new framework to quantitatively understand collective patch foraging dynamics in a hierarchical group. Through analytical formal analysis and numerical simulations, we show that it is possible to characterize how different information sharing mechanisms lead to a variety of emergent collective dynamics. In particular, we found that being a following individual can be beneficial (i.e. increase accuracy) when agents count how many leaders are in their patch (counting coupling) or observe leader arrivals in their patch (arrival pulsatile coupling). Another result is that although misinformation from leading individuals is detrimental to the following group, other positive effects can emerge, such as greater cohesion. In fact, more cohesion can be interesting in the presence of threats, as it provides risk dilution~\cite{siegfriedFlockingAntipredatorStrategy1975}\cite{powellExperimentalAnalysisSocial1974}. Non-cooperative behaviors such as misinformation or sabotage (increased travel time, see~\cite{blummoyseSocialPatchForaging2024}) could be especially advantageous for agents in case of competition for resources, e.g. depleting patches.\\
The diversity and non-trivial emergent processes found in our study highlight the mechanistic and flexible features of our model, which makes it able to understand the large diversity of animal behaviors and uncover their hidden mechanisms. In particular, social hierarchy changes the collective dynamics compared to an egalitarian group~\cite{blummoyseSocialPatchForaging2024}. For example, we found a decreased accuracy with diffusive coupling in a hierarchical group, whereas no change was observed in an egalitarian organization. In addition, different normalization conditions, by the number of agents in a patch $n_L^k(t)$, instead of the group size $N_L$, may also change the emergent properties. i.e. whether agents keep in mind the size of the L group, without observing them, or adjust their behavior depending on the observable number of L agents. For example, accuracy decreases or increases with departure pulsatile coupling, depending on whether the reference is $N_L$ or $n_L^k(t)$ (see supplementary Fig.~\ref{SuppFig}).\\ Our study also suggests that although a widely used metric in foraging studies is patch residence time (a key variable in the Marginal Value Theorem~\cite{charnovOptimalForagingMarginal1976}) accuracy and cohesion may be of critical importance to characterize group dynamics (as shown in some experimental studies~\cite{franksSpeedCohesionTradeoffs2013}\cite{stroeymeytImprovingDecisionSpeed2010}). As for pulsatile or diffusive coupling, the non-trivial variations of social cohesion strongly point out this metric to be a key descriptor to identify precisely underlying cognitive mechanisms of emergent group processes.\\
We may note that for some couplings, like pulsatile, large L groups were mandatory to have a theory-simulation match. The results may be different for smaller groups, especially for cohesion which is more variable compared to equilibrium metrics (see supplementary Fig.~\ref{SuppFig}(a)). Further directions include investigations of how different L dynamics (for example different cohesion features of the L group) may lead to different emergent processes in the F group. In addition, 
variations of existing couplings would be studied, such as repulsive counting and pulsatile coupling, as well as adding uncertainty, such as imprecise counting. Lastly, a gradual hierarchy with more than two groups would advance our understanding of the wide dynamic range of social organizations.\\
Our aforementioned results and our mechanistic modeling framework (which has been applied in a variety of cases beforehand~\cite{davidsonForagingEvidenceAccumulation2019}\cite{bidariStochasticDynamicsSocial2022}\cite{blummoyseSocialPatchForaging2024}) call for the reformulation of optimal foraging theories in a manner that accounts for the diversity of social dynamics. Our quantitative approach provides the foundations for a reformulated foraging theory.\\

\paragraph*{Acknowledgments---} We would like to thank Jacob Davidson and Zachary Kilpatrick  for insightiful discussions.  All authors acknowledge support from the Deutsche Forschungsgemeinschaft (DFG, German Research Foundation) under Germany’s Excellence Strategy-EXC 2117–422037984.\\

Authors Lisa Blum Moyse and Ahmed El Hady have developed the study, derived the results and wrote the manuscript.



\section*{Appendix}
\label{appendix}
\textbf{Calculation of $\bigl<x_L\bigr>$}\\

The leading averaged decision variable $\bigl<x_L\bigr>(t)$ is found through the average with the probability to get $x$ at a time $t$, $U^k(x,t)$. So that $\bigl<x_L\bigr>(t) = \int_\theta^0 x U^k(x,t) dx$.\\ 

The probability $U^k(x,t)$ is itself calculated through the convolution between the unit $x$ probability for a single patch problem $u(x,t)$~\cite{coxTheoryStochasticProcesses1977}(see eq.~\ref{uxl}) and the arriving distribution from the patch $k'$ to $k$, $P^{k'}(t)$. So that $U^k(x_L,t) = \int_0^t u^k(x_L,t-\tau)P^{k'}(\tau)d\tau$, with 
\begin{equation}
\begin{split}
    u(x,t) = &\frac{1}{\sqrt{4\pi Bt}}\Biggl(\exp\biggl(-\frac{(x-\alpha t)^2}{4Bt}\biggr) \\&- \exp\biggl(-\frac{\alpha\theta}{B} - \frac{(x+2\theta-\alpha t)^2}{4Bt}\biggr)  \Biggr)   
\end{split}
\label{uxl}
\end{equation} \\

\textbf{Results for normalization by $n_L^k$ instead of $N_L$}\\

In case of couplings normalized by the number of leading agents in a patch $k$, $n^k_L$, instead of the size of the leading group $N_L$, the effective drift terms are, for:\\

{\flushleft Reward coupling: $\widetilde{\alpha}_F^k(t)  = \alpha - p^k -\kappa_r \zeta^k$}\\
Pulsatile coupling: $\widetilde{\alpha}_F^k(t) = \alpha -p^k + \kappa_d \frac{P^{k}_L(t)}{Q^{k}_L(t)} - \kappa_a \frac{P^{k'}_L(t)}{Q^{k'}_L(t)}$\\
Diffusive coupling: $\widetilde{\alpha}^k_F(t) = \alpha - p^k - \kappa_\text{diff}^k \frac{\bigl<x_L^k\bigr>}{Q^k_L}\bigl(t\bigr)$\\

These effective drift terms are then used to compute the different distributions and metrics.
See supplementary Fig.~\ref{SuppFig} for the dynamics of these couplings normalized by $n_L^k$.\\




\bibliography{SocialForaging}

\renewcommand{\theHfigure}{S\arabic{figure}}
\renewcommand{\thefigure}{S\arabic{figure}}
\setcounter{figure}{0}

\begin{figure*}
    \centering
\includegraphics[width=0.95\linewidth]{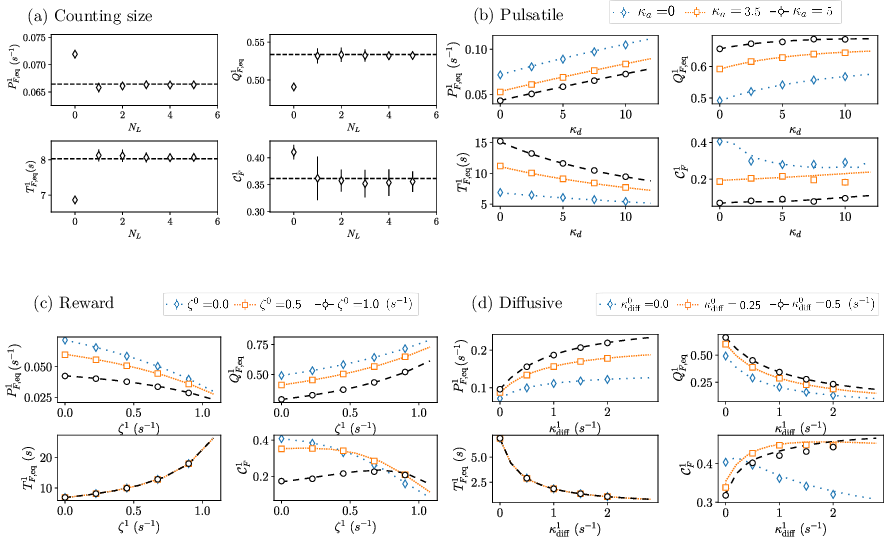}
    \caption{Effect of L group size and normalization by $n^k_L$. Best patch metrics: equilibrium leaving time distribution $P_{F,\text{eq}}^1$, equilibrium accuracy $Q_{F,\text{eq}}^1$, equilibrium patch residence time $T_{F,\text{eq}}^1$, and cohesion $\mathcal{C}_F^1$. (a) Effect of L group size $N_L$ for theory - numerical simulations match in counting coupling. Parameters are $\kappa_c=0.2$~s$^{-1}$, $\eta = 0.25$ $N_F=200$, 20 simulations, $dt = 0.001$~s. (b) Pulsatile coupling metrics as a function of the departure strength parameter $\kappa_d$ for different arrival strength parameter $\kappa_a$ values. Parameters are $N_L=60000$, $N_F=2000$, $2$ simulations, $dt=0.02$~s. (c) Reward coupling metrics as a function of the communicated reward rates in patch 1 $\zeta^1$, for different communicated reward rates in patch 0 $\zeta^0$ values. Parameters are $\kappa_r=0.5$, $N_L=2000$, $N_F=2000$, $2$ simulations, $dt=0.02$~s. (d)
    Diffusive coupling metrics as a function of the diffusive strength parameter in patch 1 $\kappa_\text{diff}^1$, for different diffusive strength parameter in patch 0 $\kappa_\text{diff}^0$ values. Parameters are $N_L = 10000$, $N_F = 2000$, 2 simulations, $dt = 0.01$~s. Lines correspond to theoretical predictions and dots to numerical simulations. Error bars correspond to the standard deviation.}
    \label{SuppFig}
\end{figure*}


\end{document}